\documentclass[11pt, oneside]{article} 

\usepackage{mathtools}
\DeclarePairedDelimiter\bra{\langle}{\rvert}
\DeclarePairedDelimiter\ket{\lvert}{\rangle}
\DeclarePairedDelimiterX\braket[2]{\langle}{\rangle}{#1\,\delimsize\vert\,\mathopen{}#2}

\usepackage{geometry} 
\usepackage{graphicx} 

\usepackage{amssymb}

\title{\huge A Quadratic Sample Complexity Reduction for Agnostic Learning via Quantum Algorithms}
\author{Daniel Z. Zanger\\danielzanger@gmail.com}

\date{} 

\begin{document}

\maketitle
\noindent \bf Abstract:\rm\,Using quantum algorithms, we obtain, for accuracy $\epsilon>0$ and confidence $1-\delta,0<\delta<1,$ a new sample complexity upper bound of $O((\mbox{\rm log}(\frac{1}{\delta}))/\epsilon)$ as $\epsilon,\delta\rightarrow 0$ for a general agnostic learning model, provided the hypothesis class is of finite cardinality. This greatly improves upon a corresponding sample complexity of asymptotic order $\Theta((\mbox{\rm log}(\frac{1}{\delta}))/\epsilon^{2})$ known in the literature to be attainable by means of classical (non-quantum) algorithms for an agnostic learning problem also with hypothesis set of finite cardinality (see, for example, Arunachalam and de Wolf (2018) and the classical statistical learning theory references cited there). Thus, for general agnostic learning, the quantum speedup in the rate of learning that we achieve with respect to these results is quadratic in $\epsilon^{-1}$.


\section{Introduction}
In classical learning theory, a principal, general result (see, for example, Lee (1996)) asserts that $N=O((\mbox{\rm log}(\frac{1}{\delta})+\mbox{\rm log}(\frac{1}{\epsilon}))/\epsilon^2)$ random (i.i.d.) samples as $\epsilon,\delta \rightarrow 0$ are sufficient to
 compute, in the setting of an agnostic learning model (see Lee et al. (1998), Arunachalam and de Wolf (2018), or \S 2 here), a function in a hypothesis class ${\cal{H}}$ (which we also call a hypothesis set) minimizing the approximation error of learning up to accuracy $\epsilon >0$
 and confidence $1-\delta,0<\delta <1,$ given a uniformly bounded squared loss function and ${\cal{H}}$ of finite VC-dimension
 (pseudo-dimension). Hence, in other words, $N=O((\mbox{\rm log}(\frac{1}
{\delta})+\mbox{\rm log}(\frac{1}{\epsilon}))/\epsilon^2)$ may be interpreted as an upper bound on the corresponding classical sample complexity for this problem. 
In this paper, we show that certain quantum learning algorithms provide a general and very significant improvement with respect to this sample complexity upper bound,
at least for the case of a hypothesis class ${\cal{H}}$ of finite cardinality.
Indeed, we show that, within the general context of an agnostic learning model and for ${\cal{H}}$ with cardinality $|{\cal{H}}|<\infty$, we can improve the cited upper bound on sample complexity from $N=O((\mbox{\rm log}(\frac{1}{\delta})+\mbox{\rm log}(\frac{1}{\epsilon}))/\epsilon^2)$ to $N=O((\mbox{\rm log}(\frac{1}
{\delta}))/\epsilon).$ 

Enabling us to achieve this very general, super-quadratic (with respect to the accuracy parameter $\epsilon$) reduction in sample complexity within this context (relative to the cited result in Lee(1996), which addresses hypothesis classes with real-valued functions) are powerful quantum algorithms, as presented in Montanaro (2015), which exploit the by now classical technique of Quantum Amplitude Estimation. These quantum algorithms obtain, up to given accuracy $\epsilon, 0<\epsilon<1,$  and (probabilistic) confidence $1-\delta,0<\delta<1,$ a dramatic asymptotic reduction in the number of samples needed to estimate the expectation of a random variable. Relying on these results, our main Theorem 4 in \S 3.2 here (also see Remark 4 in \S 3.2) establishes, given a finite hypothesis class and a uniformly bounded, nonnegative loss function, an $O((\mbox{\rm log}(\frac{1}{\delta}))/\epsilon)$-upper bound, new in the literature, on the sample complexity of learning in the case of an agnostic (quantum) learning model. This very significantly improves on an asymptotic sample complexity known to be of order $N=\Theta((\mbox{\rm log}(\frac{1}{\delta}))/\epsilon^{2})$ as $\epsilon,\delta \rightarrow 0$ for a key classical (non-quantum) agnostic learning problem (see, for example, (2) in \S 1.1.2 in Arunachalam and de Wolf (2018) -- in which the functions in the hypothesis class are assumed binary-valued -- or in fact Theorem 1 in \S 2 here). Thus, the result in Arunachalam and de Wolf (2018) establishes that, for
classical agnostic learning, $N=\Omega((\mbox{\rm log}(\frac{1}{\delta}))/\epsilon^{2})$ as $\epsilon,\delta \rightarrow 0$ random samples are asymptotically required with respect to $\epsilon$ and $\delta$, whereas we show by comparison that, for the corresponding quantum agnostic learning model we consider, just $O(\mbox{\rm log}(\frac{1}{\delta}))/\epsilon)$ samples are needed, yielding a quadratic improvement in this case (again relative to
 $\epsilon^{-1}$).

The rest of the paper is organized as follows. In \S 2 we present some underlying (classical) statistical learning theory definitions, concepts, and prior results that are useful to us in the context of the paper. 
In \S 3.1 we suitably define the concepts of Quantum Agnostic Learnability and Quantum Sample Complexity as well as state some prior results in the literature which will enable us in turn to state and establish our new upper bounds on Quantum Sample Complexity for a hypothesis set of finite cardinality under an agnostic learning model. Presented and proved in \S 3.2, these include the main, general result of the paper, Theorem 4, as well as its Corollary 2, in which Theorem 4 is specialized to the case of the standard squared loss function.

\section{Background on Statistical Learning Theory and Classical Sample Complexity}

Some of the key aims of this section are to set some basic notation and frame some crucial concepts from 
classical statistical learning theory that we will require in this section as well as the rest of the paper. 
Note that parts of our exposition here in this section somewhat
loosely follow that in Cucker and Smale (2001). 

Let ${{\cal X}}$ be a domain (i.e., connected, open subset) in, or more generally any Borel subset of, some 
$d$-dimensional Euclidean space $\mbox{\boldmath $R$}^{d}$. We denote by
$\rho$ a Borel probability measure on
$\Omega={{\cal X}}\times{{\cal Y}}$, where ${{\cal Y}}$ is a Borel subset of $\mbox{\boldmath $R$}$ which in addition
we may assume in some cases to be a bounded subset if needed. Let $z=(x,y)$ be a generic element of $\Omega.$
Write $\rho_{x}=\rho(\cdot |x)$ for the conditional probability measure on ${\cal{Y}}$ given $x\in {\cal{X}}$ induced by
$\rho$, and set
\begin{equation}
f_{\rho}(x)=\int_{\cal{Y}}y\,d\rho_{x}(y),
\end{equation}
which is the \it regression function\rm\, of $\rho$ evaluated at $x$ (assuming that the value on the right in (1) is a well-defined, finite 
integral for $x\in {\cal{X}}$). Also let $\rho_{\cal{X}}$ be the marginal probability measure of $\rho$ on $\cal{X},$ i.e., the measure on
$\cal{X}$ defined via $\rho_{\cal{X}}({\cal{X'}})=\rho(\pi^{-1}(\cal{X'}))$ for suitably measurable
$\cal{X'}\subseteq \cal{X}$, with $\pi:\Omega\rightarrow
\cal{X}$ being the projection onto the first coordinate. Similarly, let $\rho_{\cal{Y}}$ be the analogously-defined marginal probability measure of $\rho$ onto the second coordinate corresponding to ${\cal{Y}}$. We assume that $f_{\rho}\in L^{2}(\rho_{\cal{X}})= L^{2}({\cal{X}}, \rho_{\cal{X}}).$

An agnostic model of learning (see, for example, Lee (1996), Lee et al. (1998), or Arunachalam and de Wolf (2018)) is the underlying learning paradigm we adopt in this paper for the data samples.
Let $N$ denote a positive integer. Under an agnostic learning model, we consider as given a sequence 
\begin{equation}
Z_{1}=(X_{1},Y_{1}),...,Z_{N}=(X_{N},Y_{N})
\end{equation}
of $N$ independent copies of the random vector $Z=(X,Y)$ drawn from
$\Omega={\cal{X}}\times{\cal{Y}}$ according to the measure $\rho$ on
$\Omega$ as above. 
Note in particular that, within the agnostic learning paradigm, we do not
make the assumption that there is some known, fixed class of functions on ${\cal{X}}$ such that we necessarily must have $Y=g(X)$ for some function $g$ from within this class. 
We write a corresponding generic sequence of $N$ fixed members from $\Omega$ analogous to (2) as $z_{1}=(x_{1},y_{1}),...,z_{N}=(x_{N},y_{N}).$

Now let  ${\cal{H}}$ be, quite generally, any given set of (suitably measurable) real-valued functions $f:{\cal{X}}\rightarrow\mbox{\boldmath $R$}$.
We call ${\cal{H}}$ a hypothesis class or hypothesis set, and we assume that ${\cal{H}}\subseteq L^{2}({\cal{X}},\rho_{\cal{X}})$. We let $|{\cal{H}}|$ denote the cardinality of ${\cal{H}}.$
The hypothesis class  ${\cal{H}}$ is the set from which we choose a function $f_{N}$, based on a set of data samples, that
approximates the regression function $f_{\rho}$ in some suitable way.
Moreover, under the agnostic model as we frame it here, we impose a so-called proper learning requirement, by which we mean, throughout this article, that the function $f_{N}$ we seek must satisfy $f_{N}\in{\cal{H}}$. The proper learning requirement is very common in the statistical learning literature, and, as we only consider hypothesis classes of finite cardinality in this article, such a specification appears to be an especially natural one to impose.

In the case of classical learning (that is, in this case, machine learning performed by classical algorithms and in the setting of a classical data sample model as described in this section), 
it is known that, even in cases of hypothesis sets of infinite cardinality, certain additional assumptions can be imposed in the context of agnostic learning as formulated here in order to improve the rate of learning to $N=\tilde{O}(\mbox{\rm log}(\frac{1}{\delta})/\epsilon)$ from $N=\tilde{O}(\mbox{\rm log}(\frac{1}
{\delta})/\epsilon^{2})$  (where, here, $\tilde{O}(\mbox{\rm log}(\frac{1}
{\delta})/\epsilon)$ denotes an asymptotic rate of $O(\mbox{\rm log}(\frac{1}
{\delta})/\epsilon)$ up to logarithmic factors in $\epsilon$ and analogously for $\tilde{O}(\mbox{\rm log}(\frac{1}
{\delta})/\epsilon^{2})$) at least in the case of a uniformly bounded squared loss function along with a hypothesis set with finite VC-dimension (Lee (1996), Lee et al. (1998)). Such assumptions include that the regression function $f_{\rho}$ satisfy $f_{\rho}\in{\cal{H}},$ that there be some known, fixed class of functions on
${{\cal X}}$ such that we necessarily must have $Y=g(X)$ for some function $g$ from this class, or that ${\cal{H}}$ be a convex set of functions.
We note that in this paper we do not make any of these assumptions for our results on quantum learning and quantum sample complexity as we specify these terms here (in \S 3).

To measure closeness of approximation in the learning context, we also require a loss function 
\begin{equation}
L:{\cal{H}}\times\Omega\rightarrow\mbox{\boldmath $R_{+}$},\,(f,(x,y))\mapsto L(f,(x,y)),
\end{equation}
where, for us, $\mbox{\boldmath $R_{+}$}$ denotes the set of nonnegative real numbers. We furthermore define, for any given $L(\cdot,(\cdot,\cdot))$ as in (3) and $f\in{\cal{H}},$
\begin{equation}
L_{f}:\Omega\rightarrow\mbox{\boldmath $R_{+}$},\,(x,y)\mapsto L_{f}(x,y)=L(f,(x,y)).
\end{equation}
In this paper, we will simply assume that the loss function $L(\cdot,(\cdot,\cdot))$ is uniformly bounded so that 
$L(f,(x,y))\leq C$ for all $x\in{\cal{X}},\,y\in{\cal{Y}}\,$ and $f\in {\cal{H}}$ as well as some $C,0<C<\infty.$ We note, however, that, since for the main results of this paper
we only consider ${\cal{H}}$ of finite cardinality in any case, uniform boundedness here follows immediately from assuming boundedness of the function
$L(f,(\cdot,\cdot))$ separately for each function $f\in {\cal{H}}.$
One of the most common and simplest loss functions is the 0-1 loss,
with respect to which one simply takes $L(f,(x,y))=1$ if $f(x)\neq y$ and $0$ otherwise. Another common loss function is the standard squared (or squared error) loss function -- also called the quadratic loss function -- defined via 
$L(f,(x,y))=(f(x)-y)^{2}$. 

Next we define the concept of classical learnability itself. We say that a hypothesis set ${\cal{H}}$ is \it\,classically learnable\rm\, (with respect to some given loss function $L(\cdot,(\cdot,\cdot))$ as in (3)) provided there is
an integer-valued function $m(\cdot,\cdot)$ and a classical algorithm such that, if the algorithm
draws $N\geq m(\epsilon,\delta)$ i.i.d. samples as in (2) for any confidence $1-\delta,0<\delta<1,$ and accuracy $\epsilon,0<\epsilon <1,$ values, then it outputs a function $f_{N}\in {\cal{H}}$ for which
\begin{equation}
\mbox{\rm\bf P}(\mbox{\rm\bf E}[L(f_{N},(X,Y))]\leq \mbox{\rm min}_{f\in {\cal{H}}} \mbox{\rm\bf E}[L(f,(X,Y))] +\epsilon)\geq 1-\delta.
\end{equation}
Moreover, since we are presupposing in this paper an agnostic model for the data samples as described above in this section, we can refer, not only to the concept of classical learnability as just specified, but also to classical agnostic learnability.

In (5), the probability $\mbox{\rm\bf P}$ with respect to which the confidence is assessed is taken with respect to the product measure $\rho^{N}$. However, 
the expectations that appear in (5) on both the left-hand and right-hand side of the $\leq$-inequality, we note, are 
taken with respect to the random vector $(X,Y)$ (as introduced just after (2)) alone, so that, for example, we have, with respect to the expectation on the left in (5),
\begin{equation}
\mbox{\rm\bf E}[L(f_{N},(X,Y))]=\int_{\Omega}L(f_{N},(x,y))d\rho (x,y).
\end{equation}
We call the quantity $\mbox{\rm min}_{f\in {\cal{H}}}\mbox{\rm\bf E}[L(f,(X,Y))]$  appearing in (5) the approximation error
(with respect to ${\cal{H}}$). 

We also note here that, for some loss functions or types of distributions for the samples $(X,Y)$, a fast learning rate of $N=\tilde{O}((\mbox{\rm log}(\frac{1}
{\delta}))/\epsilon)$ may be achieved as well by inserting an additional $(1+r)$-factor, $r>0,$ into (5) and considering 
\begin{equation}
\mbox{\rm\bf P}(\mbox{\rm\bf E}[L(f_{N},(X,Y))]\leq (1+r)\mbox{\rm min}_{f\in {\cal{H}}} \mbox{\rm\bf E}[L(f,(X,Y))] +\epsilon)\geq 1-\delta
\end{equation}
instead. The expression in (7) is closely related to the notion of nonexact oracle inequalities (see Lecue and Mendelson (2012)) whereas (5) is associated, by comparison, with the concept of an exact oracle inequality. Of course,
if ${\rm min}_{f\in {\cal{H}}} \mbox{\rm\bf E}[L(f,(X,Y))]>0,$ then the case $r=0$ in (7) (that is, the case of (5)) is preferred as an approximation as close as possible to 
the exact quantity ${\rm min}_{f\in {\cal{H}}} \mbox{\rm\bf E}[L(f,(X,Y))]$ is generally desired.

The\,\it classical sample complexity\,\,\rm of learning a hypothesis class ${\cal{H}}$ we now define to be the function $m(\cdot,\cdot)$ with the smallest value for $m(\epsilon,\delta)$ for each pair
$(\epsilon,\delta)$ such that an algorithm for learning ${\cal{H}}$ as in (5) above exists. Theorem 1 below gives optimal upper and lower bounds on the classical sample complexity for agnostic learning in a suitable context.

\noindent\bf Theorem 1.\rm (see Arunachalam and de Wolf (2018)) \rm\it For any given positive integer $d,$ let ${\cal{X}}=\{0,1\}^{d},$ ${\cal{Y}}=\{0,1\},$ and define the hypothesis class ${\cal{H}}$ to 
be the (finite) set of all functions $f$ mapping $\{0,1\}^{d}$ into $\{0,1\}.$  Then, for $\epsilon >0$ and $\delta,0<\delta<1,$ 
\begin{equation}
N=\Theta \left(\frac{\mbox{\rm log}(\frac{1}{\delta})}{\epsilon^2}\right)\,\,\mbox{\rm as}\,\,\epsilon,\delta\rightarrow 0
\end{equation}
random samples, as in (2), are both necessary and sufficient to compute a function $f_{N}\in {\cal{H}}$ for which
\begin{equation}
\mbox{\rm\bf P}\left( \mbox{\rm\bf E}[(f_{N}(X)-Y)^{2}]\leq \mbox{\rm min}_{f\in
{\cal{H}}}\mbox{\rm\bf E}[(f(X)-Y)^{2}]+\epsilon\right)\geq 1-\delta.
\end{equation}
\rm
For Theorem 1, see Arunachalam and de Wolf (2018) (in particular, see (2) in \S 1.1.2 in Arunachalam and de Wolf (2018) and the further references mentioned there in connection with this bound). The classical learning bound stated in Theorem 1 above, already known in the literature, is mainly useful here in order to compare it with corresponding sample complexity bounds, as in \S 3 below, for which quantum algorithms are utilized.

\section{Bounds on the Sample Complexity of Quantum Learning}
\subsection{Quantum Learnability and Quantum Sample Complexity}

Our initial objective in this subsection is to suitably define the concepts of Quantum Agnostic Learnability and Quantum Sample Complexity as appropriate generalizations of the corresponding classical concepts within the quantum context. All definitions formulated and notations defined in \S 2 still apply here in \S 3.1. We note as well that for background on some of the basic, general notations and underlying concepts from quantum computation appearing here see Nielsen and Chuang (2010).

We assume here that (all components of) the random vector $Z=(X,Y)$ -- as well as of course all of its copies including, for $i=1,...,N,$ the $Z_{i}=(X_{i},Y_{i})$ -- is (are) suitably approximated in some discretized way should any of its components be continuous variables. That is, we will henceforth assume that $Z=(X,Y)$ is a discrete random vector and even one taking on only finitely many possible values.

Following Montanaro (2015) (see in particular \S 2 there for further details, as well as similar and/or related treatments in Kothari and O’Donnell (2023), especially \S 2, and in \S II.A in Kaneko et al. (2021)), we now let $\cal{A}(\cdot)$ be any unitary operator (quantum circuit) satisfying, when applied to an initial input state $\ket{0^{n}},$
\begin{equation}
{\cal{A}}(\ket{0^{n}})=\sum_{z\in\{0,1\}^{k}}\alpha_{z}\ket{\psi_{z}}\ket{z}
\end{equation}
 for some normalized states $\ket{\psi_{z}}.$ Also, here in (10), the positive integer $k$ (recalling that we now assume the random vector $Z$ to be discrete as mentioned in the preceding paragraph) is presumed large enough to represent all vectors $z\in\Omega$ for which $\mbox{\rm\bf P}(Z=z)>0$ in the corresponding $2^{k}$-dimensional computational basis, and we have $n\geq k$ for the positive integer $n$. In addition, again with respect to (10), we have $|\alpha_{z}|=\sqrt{\mbox{\rm\bf P}(Z=z)}$ for each $z\in\{0,1\}^{k}$. We note as well that we will simply use the same notation $z$ to represent this vector in the computational basis as we do to denote it as a member of $\Omega,$ so that, for $f\in{\cal{H}}$ and $z\in\{0,1\}^{k},\,L_{f}(z)$ is defined to be its value for $z$ considered as a member of $\Omega$.

We now can, assuming access to both the unitary circuit ${\cal{A}}$ as in (10) as well as its inverse ${\cal{A}}^{-1}$, define a \it\,quantum sample\rm\, to be a single call, as defined by (10), to the quantum circuit ${\cal{A}}(\cdot)$. That is, a quantum sample is a measurement of the state ${\cal{A}}(\ket{0^{n}})$ (as in (10)). Note as well that such a call to ${\cal{A}}$ may be viewed as corresponding to a sampling from the random vector 
$Z=(X,Y)$ 
due to the fact that, considering (10), $|\alpha_{z}|^{2}=\mbox{\rm\bf P}(Z=z)$ for $z\in \{0,1\}^{k}.$ This definition of a quantum sample specifies the (quantum) data input model for learning that we presuppose for the principal results of this paper.

Let $L(\cdot,(\cdot,\cdot))$ be a given loss function as in (3). We say that the hypothesis set ${\cal{H}}$ is\it\,quantum learnable\rm\,(with respect to this loss function) provided there is an integer-valued function $m(\cdot,\cdot)$ and an algorithm such that, if the algorithm
executes $N\geq m(\epsilon,\delta)$ calls to ${\cal{A}}(\ket{0^{n}})$ as in (10) for any confidence $1-\delta,0<\delta<1,$ and accuracy $\epsilon,0<\epsilon <1$, values (that is, if $N\geq m(\epsilon,\delta)$ quantum samples as defined in the previous paragraph are inputted into the algorithm), then it outputs a function $f_{N}\in {\cal{H}}$ for which
\begin{equation}
\mbox{\rm\bf P}(\mbox{\rm\bf E}[L(f_{N},(X,Y))]\leq \mbox{\rm min}_{f\in {\cal{H}}} \mbox{\rm\bf E}[L(f,(X,Y))] +\epsilon)\geq 1-\delta.
\end{equation}
If ${\cal{H}}$ is quantum learnable in the sense just defined then we say that the algorithm quantum learns ${\cal{H}}.$ The \it\,quantum sample complexity\rm\, of (quantum) learning ${\cal{H}}$ is the function with the smallest value for $m(\epsilon,\delta)$ for each pair $(\epsilon,\delta)$ such that an algorithm for quantum learning ${\cal{H}}$ as in (11) above exists. If samples as in (2) drawn from the (classical) random vector $Z=(X,Y)$ are consistent with (the generality of) an agnostic learning model as described in \S 2, which we do presuppose them to be in this paper, then we say in this context that, furthermore, ${\cal{H}}$ is quantum agnostic learnable, and we can speak of the concept of quantum agnostic learnability and also that of the quantum sample complexity of agnostic learning or, equivalently, the sample complexity of quantum agnostic learning.

\noindent\bf Remark 1.\rm As discussed in the Introduction, our results here establish, for the sample complexity of quantum agnostic learning, a quadratic reduction (with respect to the accuracy parameter $\epsilon$) relative to the best known bounds for classical agnostic learning sample complexity and, in some important cases at least, relative to the best possible bounds for classical agnostic learning sample complexity (see, for example, Theorem 1 in \S 2 here). Indeed, Arunachalam and de Wolf (2018) obtain quantum sample complexity bounds of $N=\Theta((\mbox{\rm log}(\frac{1}{\delta}))/\epsilon^{2})$ for the asymptotic number of quantum samples (according to their definition) both necessary and sufficient to learn an unknown distribution under the quantum agnostic model of learning that they consider. Thus their results imply that no sample complexity reduction (other than possibly with respect to smaller constant factors) can be achieved relative to the case of classical agnostic learning for the quantum agnostic learning model that they analyze. The reason for this apparent discrepancy in comparison with our own results here appears to involve the type of quantum samples or examples considered in the quantum agnostic learning model studied by Arunachalam and de Wolf (2018), as well as how much information the learner has concerning how these samples are being generated.  Arunachalam and de Wolf (2018) assumes that the sample data available to the learner are quantum examples that are (coherent) quantum states of the form
\begin{equation}
\sum_{z\in\{0,1\}^{m}}\sqrt{p_{z}}\ket{z}
\end{equation}
for suitable positive integers $m$ and where $p_{z}=\mbox{\bf P}(Z=z)$. On the other hand, the quantum data input model presupposed in this paper is somewhat different. Note first that quantum samples  (see (10)) as defined for the results in this paper, as well as those in Montanaro (2015), are not assumed to be coherent states as they are in Arunachalam and de Wolf (2018) (see (12)). Also however, presupposing in this paper the same (quantum) data input model as is employed in Montanaro (2015) (see \S 2 there), we assume that the learner has access not just to the (quantum) data samples themselves as specified by (10) but also to the unitary quantum circuit ${\cal{A}}$ which generates them as well as to its inverse ${\cal{A}}^{-1}$.  This is so the learner can, as prescribed by Quantum Amplitude Estimation (restated as Theorem 2 here), implement the operator ${\cal{U}}=2\ket{\psi}\bra{\psi}-I$ as well (in this connection, see also Theorem 3 here as well as our Remark 3 below). For these reasons we can say that the results in Arunachalam and de Wolf (2018) are not inconsistent with our own results here, concluding Remark 1.

Our next objective is to present Algorithm 1, a quantum algorithm which will enable us to achieve the advertised quadratic (with respect to the accuracy parameter $\epsilon$) reduction in sample complexity for agnostic learning. In order to formulate it, we well need the following by now classical result concerning Quantum Amplitude Estimation.

\noindent\bf Theorem 2.\rm (Theorem 2.2 as stated in Montanaro (2015))\it\,There is a quantum algorithm called amplitude
estimation which takes as input one copy of a quantum state $\ket{\psi}$, a unitary transformation ${\cal{U}}=2\ket{\psi}\bra{\psi}-I,$ a unitary transformation $V=I-2P$ for some projector $P,$ and an integer $t$. The algorithm outputs $\tilde{a}$, an estimate of $a=<\psi|P|\psi>,$ such that
\begin{equation}
|\tilde{a}-a|\leq \frac{2\pi\sqrt{a(1-a)}}{t}+\frac{\pi^{2}}{t^{2}}
\end{equation}
with probability at least $\frac{8}{\pi^{2}}$, using ${\cal{U}}$ and $V$ $t$ times each.\rm

Algorithm 1, which we now state, is essentially Algorithm 1 in \S 2(a) of Montanaro (2015). 

\begin{center}
\bf\large Algorithm 1
\end{center}
\noindent----------------------------------------------------------------------------------------------------------------------

\noindent\bf Input:\rm\, A unitary operator ${\cal{A}}$ as in (10), a uniformly bounded loss function $L(\cdot,(\cdot,\cdot))$ as in (3), 
a function $f\in{\cal{H}}$ (with ${\cal{H}}$ being the chosen hypothesis class), an integer $t$, a real value $\delta>0,$ 
and positive integers $k$ and $n$ (where $k$ and $n$ are as described just after (10) above).

(i) Let $W$ be the unitary operator on $k+1$ qubits defined by 
\begin{equation}
W\ket{z}\ket{0}=\ket{z}(\sqrt{1-L_{f}(z)}\ket{0}+\sqrt{L_{f}(z)}\ket{1}),\,z\in\{0,1\}^{k}.
\end{equation}

(ii) Repeat the following step $\Omega(\mbox{\rm log}(\delta^{-1}))$ times and output the median of the 

\,\,\,\,\,\,\,\,\,\,results:

\,\,\,\,\,\,\,\,\,\,\,\,\,\,\,\,(a)\,Apply $t$ iterations of Amplitude Estimation as stated in Theorem 2 above, 

\,\,\,\,\,\,\,\,\,\,\,\,\,\,\,\,\,\,\,\,\,\,\,\,\,setting $\ket{\psi}=(I\otimes W)({\cal{A}}\otimes I)\ket{0^{(n+1)}}$ and $P=I\otimes\ket{1}\bra{1}.$

\noindent---------------------------------------------------------------------------------------------------------------------

\noindent\bf Remark 2.\rm We note that Algorithm 1 as formulated in Montanaro (2015) would require, with respect to our own Algorithm 1 as stated above, the condition $0\leq L_{f}(z)\leq 1$ for all $z=(x,y)\in\Omega$ and any $f\in {\cal{H}}$. Consider that, because we assume for our results here that the loss function $L_{f}(z)=L(f,z)$, for all $z=(x,y)\in\Omega, f\in {\cal{H}}$, is uniformly bounded as well as nonnegative, multiplication of $L_{f}(z)$ 
by a suitable positive constant will imply that the condition $0\leq L_{f}(z)\leq 1$
is satisfied yet does not affect our ability to obtain the results we seek. Therefore, we can safely ignore the condition $0\leq L(\cdot,\cdot)\leq 1$ appearing in Algorithm 1 in Montanaro (2015), concluding Remark 2.

Before stating Theorem 3 below, we note here that, with respect to its statement as well as within the rest of the article, the constant $C,0<C<\infty,$ may denote different values at different locations within the paper (including, for example, different locations within the statement of a single theorem or corollary).

\noindent\bf Theorem 3.\rm (Montanaro (2015)) \rm\it  Let $L(\cdot,(\cdot,\cdot))$ be any given loss function, as defined by (3), for which $L(f,(x,y))\leq C$ for all $x\in{\cal{X}}$ and  $y\in{\cal{Y}}$ as well as some given $f\in {\cal{H}}$ and some $C,0<C<\infty.$ Let as well $\ket{\psi}$ be defined as in Algorithm 1, and set ${\cal{U}}=2\ket{\psi}\bra{\psi}-I$. Given any numbers $\delta,0<\delta <1,$ and $\epsilon,0<\epsilon<1,$\,$O(\mbox{\rm log}(\delta^{-1}))$ copies of the state
${\cal{A}}(\ket{0^{n}})$ as in (10) and $O\left(\frac{\mbox{\rm log}(\delta^{-1})}{\epsilon}\right)$ uses of ${\cal{U}}$ are sufficient, using Algorithm 1, to produce an estimate $\tilde{\mu}$ for $\mbox{\rm\bf E}[L_{f}(Z)]$ satisying
\begin{equation}
|\mbox{\rm\bf E}[L_{f}(Z)]-\tilde{\mu}|\leq\epsilon
\end{equation}
with probability at least $1-\delta$. \rm

Theorem 3 follows directly from Theorem 2.3 in Montanaro (2015) upon taking the random variable $\nu({\cal{A}})$ in the notation there to be defined to be $\nu({\cal{A}})=L_{f}(Z)=L_{f}(X,Y)$ for given $f\in{\cal{H}}$. To apply quantum amplitude estimation as in Theorem 2 above, note as well that, as in the proof of Theorem 2.3 in Montanaro (2015),
\begin{equation}
\mbox{\rm\bf E}[\nu({\cal{A}})]=\mbox{\rm\bf E}[L_{f}(Z)]=\sum_{z\in\{0,1\}^{k}}|\alpha_{z}|^{2}L_{f}(z)=<\psi| P | \psi>
\end{equation}
where, here, the state $\ket{\psi}$ and operator $P$ are as given in Algorithm 1 above.

\noindent\bf Remark 3.\rm\, From the definitions of ${\cal{U}}$ as in the statement of Theorem 3 and of $\ket{\psi}$ in Algorithm 1 above -- and as observed in \S 2(a) in Montanaro (2015) -- we note that ${\cal{U}}$ can be implemented by means of one use of ${\cal{A}}$ along with one use of ${\cal{A}}^{-1}$ (in this connection, see also Table 1 in Montanaro (2015)), concluding Remark 3.

Considering Remark 3, we have the following corollary of Theorem 3.  

\noindent\bf Corollary 1.\rm\it Let $L(\cdot,(\cdot,\cdot))$ be any given loss function, as defined by (3), for which $L(f,(x,y))\leq C$ for all $x\in{\cal{X}}$ and  $y\in{\cal{Y}}$ as well as some given $f\in {\cal{H}}$ and some $C,0<C<\infty.$ Also let $\epsilon,0<\epsilon <\frac{1}{4},$ and $\delta,0<\delta <\frac{1}{2},$ be any two numbers respectively from the specified intervals. Then, $N$ calls to ${\cal{A}}(\ket{0^{n}})$ as in (10) -- that is, $N$ quantum samples (as defined above within this subsection) -- are sufficient, using Algorithm 1,
to produce an estimate $\tilde{\mu}=\tilde{\mu}(L,N,f)$ of $\mbox{\rm\bf E}[L_{f}(X,Y)]$ such that 
\begin{equation}
\mbox{\rm\bf P}\left(|\mbox{\rm\bf E}[L_{f}(X,Y)]-\tilde{\mu}(L,N,f)|\leq\epsilon\right)\geq 1-\delta,
\end{equation}
provided that
\begin{equation}
N\geq \frac{C\mbox{\rm log}(\delta^{-1})}{\epsilon}
\end{equation}
for some constant $C,0<C<\infty.$ \rm

\subsection{New Bounds on the Sample Complexity of Quantum Agnostic\\ Learning}
Our principal goal in this subsection is to state and prove Theorem 4, which is new in the literature and is the main result of the paper. Note that all definitions formulated and notations introduced in \S 2 as well as in \S 3.1
still apply here in \S 3.2.

Assume that some loss function $L(\cdot,(\cdot,\cdot))$ is given. Define, for any (finite) hypothesis set $\cal{H}$ and positive integer $N,$
\begin{equation}
f_{N}=f_{L,N}=\mbox{\rm arg min}_{f\in\cal{H}}\,\tilde{\mu}(L,N,f),
\end{equation}
where $\tilde{\mu}(\cdot,\cdot,\cdot)$ was defined within the statements of Theorem 3 and its Corollary 1 above. Also set
\begin{equation}
f_{\cal{H}}=f_{L,\cal{H}}=\mbox{\rm arg min}_{f\in\cal{H}}\mbox{\rm\bf E}[L_{f}(X,Y)].
\end{equation}

We now state the main result of the paper, our new upper bound on quantum sample complexity.

\noindent\bf Theorem 4.\rm\it\, Let ${\cal{H}}$ be a given hypothesis class, assumed to be of finite cardinality, that is, to satisfy $|{\cal{H}}|<\infty.$ Suppose as well that $L(\cdot,(\cdot,\cdot))$ is any given loss function as in (3) that is uniformly bounded in the sense that $L(f,(x,y))\leq C$ for all $x\in{\cal{X}},$ all $y\in{\cal{Y}},$ and all $f\in {\cal{H}},$ as well as some $C,0<C<\infty.$ Then, for any numbers $\epsilon,0<\epsilon<\frac{1}{8},$ and $\delta,0<\delta<\frac{1}{2},\,N$ calls to ${\cal{A}}(\ket{0^{n}})$ as in (10) -- that is, $N$ quantum samples (as defined in \S 3.1) -- are sufficient to ensure, defining the function $f_{N}\in{\cal{H}}$ as in (19), that
\begin{equation}
\mbox{\rm\bf P}\left( \mbox{\rm\bf E}[L(f_{N},(X,Y))]\leq \mbox{\rm min}_{f\in
{\cal{H}}}\mbox{\rm\bf E}[L(f,(X,Y))]+\epsilon\right)\geq 1-\delta
\end{equation}
holds, provided that
\begin{equation}
N\geq \frac{C(\mbox{\rm log}(|{\cal{H}}|)+\mbox{\rm log}(\delta^{-1}))}{\epsilon}
\end{equation}
for some constant $C,0<C<\infty.$

\noindent\bf Remark 4.\rm\, Clearly, (21)-(22) in Theorem 4 imply a (quantum) sample complexity upper bound of $O(\mbox{\rm log}(\frac{1}{\delta})/\epsilon)$ as $\epsilon,\delta\rightarrow 0$ for agnostic learning. We further note here that the hypotheses of Theorem 4 are general enough to encompass the setting of Theorem 1 in \S 2. So, we see that the sample complexity upper bound of $O(\mbox{\rm log}(\frac{1}{\delta})/\epsilon)$ established by Theorem 4 offers a quadratic reduction, with respect to the accuracy parameter $\epsilon$, relative to the corresponding optimal classical asymptotic sample complexity of order $\Theta(\mbox{\rm log}(\frac{1}{\delta})/\epsilon^{2})$ as presented in Theorem 1. Thus we see that Theorem 4 offers a quadratic quantum speedup in the rate of learning relative to the bounds of Theorem 1, concluding Remark 4.

Now, in order to prove Theorem 4, we require the following lemma.

\noindent\bf Lemma 1.\rm\it\,Let $L(\cdot,(\cdot,\cdot))$ be any given loss function as in (3), and suppose that the hypothesis class ${\cal{H}}$ is of finite cardinality. Assume as well that, for some given values $\epsilon,0<\epsilon<1,$ and $\delta,0<\delta<1,$ as well as some integer $N>0,$ we have
\begin{equation}
\mbox{\rm\bf P}\left(\mbox{\rm max}_{f\in{\cal{H}}}|\mbox{\rm\bf E}[L(f,(X,Y))]-\tilde{\mu}(L,N,f)|\leq\epsilon\right)\geq 1-\delta,
\end{equation}
where, for each $f\in{\cal{H}}$, $\tilde{\mu}(L,N,f)$ was defined in the statement of Corollary 1. Then, if $f_{N}$ is as in (19), it follows that
\begin{equation}
\mbox{\rm\bf P}\left( \mbox{\rm\bf E}[L(f_{N},(X,Y))]\leq \mbox{\rm min}_{f\in
{\cal{H}}}\mbox{\rm\bf E}[L(f,(X,Y))]+2\epsilon\right)\geq 1-\delta.
\end{equation}
\noindent\rm\bf Proof.\,\rm  Consider that (23) implies that both
\begin{equation}
\mbox{\rm\bf E}[L(f_{N},(X,Y))]\leq \tilde{\mu}(L,N,f_{N})+\epsilon
\end{equation}
and 
\begin{equation}
 \tilde{\mu}(L,N,f_{{\cal{H}}}) \leq \mbox{\rm\bf E}[L(f_{{\cal{H}}},(X,Y))]+\epsilon
\end{equation}
simultaneously hold with probability no smaller than $1-\delta$. Moreover, by definition, we have that
\begin{equation}
 \tilde{\mu}(L,N,f_{N})\leq \tilde{\mu}(L,N,f_{{\cal{H}}}).
\end{equation}
Hence, with probability at least $1-\delta,$
\begin{equation}
\mbox{\rm\bf E}[L(f_{N},(X,Y))]\leq \tilde{\mu}(L,N,f_{N})+\epsilon\leq \tilde{\mu}(L,N,f_{{\cal{H}}}) +\epsilon\leq \mbox{\rm\bf E}[L(f_{{\cal{H}}},(X,Y))]+2\epsilon.
\end{equation}
So, (24) follows.\,\,$\square$

\noindent\bf Proof of Theorem 4.\rm\,First consider that Corollary 1 implies that, for any given $f\in{\cal{H}}$ and any $\epsilon,0<\epsilon<1,$
\begin{equation}
\mbox{\rm\bf P}\left(|\mbox{\rm\bf E}[L(f,(X,Y))]- \tilde{\mu}(L,N,f)|\leq\epsilon\right)\geq 1-e^{-CN\epsilon}.
\end{equation}
Hence, elementary rules of probability imply that
\begin{equation}
\mbox{\rm\bf P}\left(\mbox{\rm max}_{f\in{\cal{H}}}|\mbox{\rm\bf E}[L(f,(X,Y))]-\tilde{\mu}(L,N,f)|\leq\epsilon\right)\geq 1-|{\cal{H}}|e^{-CN\epsilon}.
\end{equation}
Thus, taking $\delta\geq |{\cal{H}}|e^{-CN\epsilon},$ (23) in Lemma 1 is satisfied, and (24) then implies (21)-(22) in the statement of the theorem.\,\,$\square$

Finally, in Corollary 2 below, we specialize the statement of Theorem 4 to the case of the standard squared loss function. Concerning this, the following proposition will be useful.

\noindent\bf Proposition 1.\rm (Cucker and Smale (2001))\it\,Suppose that ${\cal{Y}}$ is a bounded subset of $\mbox{\boldmath $R$}$ and that $f\in{\cal{H}}$ is a bounded function. Then we have
\begin{equation}
\int_{\Omega}(f(x)-y)^{2}d\rho (x,y)=\int_{\cal{X}}(f(x)-f_{\rho}(x))^{2}\,d\rho_{\cal{X}}(x)
+\sigma_{\rho}^{2},
\end{equation}
where
\begin{equation}
\sigma_{\rho}^{2}=\int_{\Omega}(y-f_{\rho}(x))^{2}d\rho (x,y)=\int_{\cal{X}}\left(
\int_{{\cal{Y}}}(y-f_{\rho}(x))^{2}d\rho_{x}(y)\right) d\rho_{\cal{X}}(x).
\end{equation}

\noindent\bf Corollary 2.\rm\it\,Suppose that hypothesis set ${\cal{H}}$ is of finite cardinality and also is such that $|f(x)|\leq C$ for all $x\in{\cal{X}}$ and $f\in{\cal{H}},$ as well as some $C,0<C<\infty.$
Assume as well that the set ${\cal{Y}}$ is a bounded subset of $\mbox{\boldmath $R$}$.
Then, for any numbers $\epsilon,0<\epsilon<\frac{1}{8},$ and $\delta,0<\delta<\frac{1}{2},\, N$ calls to ${\cal{A}}(\ket{0^{n}})$ as in (10) -- that is, $N$ quantum samples (as defined in \S 3.1) --are sufficient to ensure that, with probability at least $1-\delta$,
\begin{equation}
\mbox{\rm\bf E}[(f_{N}(X)-Y)^{2}]\leq \mbox{\rm min}_{f\in {\cal{H}}} \mbox{\rm\bf E}[(f(X)-Y)^{2}]
+ \epsilon,
\end{equation}
where the function $f_{N}$ is as in (19), provided that
\begin{equation}
N\geq \frac{C(\mbox{\rm log}(|{\cal{H}}|)+\mbox{\rm log}(\delta^{-1}))}{\epsilon}
\end{equation}
for some constant $C,0<C<\infty.$ Furthermore, for $N$ once again satisfying (34), we also have that
\begin{equation}
\mbox{\rm\bf P}\left(|| f_{N}-f_{\rho}||^{2}_{L^{2}(\rho_{\cal{X}})}\leq \mbox{\rm min}_{f\in
{\cal{H}}}||f-f_{\rho}||^{2}_{L^{2}(\rho_{\cal{X}})}+\epsilon\right)\geq 1-\delta
\end{equation}
where $L^{2}(\rho_{\cal{X}})=L^{2}({\cal{X}},\rho_{\cal{X}})$.

\rm We note that (33)-(34) in Corollary 2 follow directly from (21)-(22) in Theorem 4. Then,
(35) follows from (33), upon applying Proposition 1 above to both sides in (33).

\begin{center}
\large\bf References
\end{center}

\noindent [1] Arunachalam, S. and de Wolf, R. (2018) Optimal quantum sample
complexity of learning algorithms. Journal of Machine Learning Research 19(71):1–36.

\noindent [2] Cucker, F. and Smale, S. (2001) On the mathematical foundations of learning. Bull.
Amer. Math. Soc. (N.S.) 39(1): 1-49.

\noindent [3] Kaneko, K., Miyamoto, K., Takeda, N., and Yoshino, K. (2021) Quantum speedup of Monte Carlo integration
with respect to the number of dimensions and its application to finance. Quantum Information Processing, 20(185). 

\noindent [4] Kothari, R. and O’Donnell, R. (2023) Mean estimation when you have the source code; or,
quantum Monte Carlo methods. Proceedings of the 2023 Annual ACM-SIAM Symposium on
Discrete Algorithms (SODA), pp. 1186–1215, SIAM.

\noindent [5] Lecue, G. and Mendelson, S. (2012) General nonexact oracle inequalities for classes with a subexponential envelope. Ann. Stat. 40,
832–860.

\noindent [6] Lee, W. S. (1996) Agnostic Learning and Single Hidden Layer Neural Networks.
PhD thesis, Dept. of Systems Engineering, Research School of Information Sciences and
Engineering, Australian National University, Canberra, Australia.

\noindent [7] Lee, W. S., Bartlett, P., and Williamson, R. C. (1998) The Importance of Convexity
in Learning with Squared Loss. IEEE Transactions on Information Theory, 44(5):1974-1980.

\noindent [8] Montanaro, A. (2015) Quantum speedup of Monte Carlo methods. Proceedings of the Royal Society A, 471: 20150301.

\noindent [9] Nielsen, M. and Chuang, I. (2010) Quantum Computation and Quantum Information: 10th Anniversary Edition,
10th ed. (Cambridge University Press, New York, NY).

\end{document}